# Low temperature magneto-dielectric coupling in nanoscale layered SmFe$_{0.5}$Co$_{0.5}$O$_3$ perovskite

Ashish Shukla[1], Akash Singh[1], Md. Motin Seikh[2] and Asish K Kundu[1*]

[1]*Discipline of Physics, Indian Institute of Information Technology, Design & Manufacturing Jabalpur, Dumna Airport Road, Madhya Pradesh–482005, India*

[2]*Department of Chemistry, Visva-Bharati, Santiniketan, West Bengal –731235, India*

**Abstract**

In this study, we determined the different physical characteristics of nanoscale layered mixed Fe-Co layers of orthoferrite SmFe$_{0.5}$Co$_{0.5}$O$_3$ based on magnetic and dielectric measurements. Magnetization analyzes showed that the system was antiferromagnetic with a magnetic transition around 310 K and two spin-reorientation transitions at around 192 K and 22 K. In this phase, the antisymmetric exchange interaction induced weak ferromagnetism due to canting of the magnetic spins in a similar manner to the Ln-Fe-Co system. Importantly, this nanoscale layered perovskite also exhibited a positive magneto-dielectric effect of around 2.5% (100 K) and the value decreased on both sides of this temperature. We analyzed the conditions related to the existence of magnetic and dielectric coupling in terms of the magnetic interactions between the cations as well as the spin–lattice interactions in the low temperature range.

*Corresponding author e-mail: asish.k@gmail.com/asish.kundu@iiitdmj.ac.in*

# 1. Introduction

Perovskite-based multiferroic materials have attracted much attention because of their wide variety of physical properties [1-5]. Designing multiferroics is a challenging issue for transition metal perovskites. The prevalence of conjugate ordering in magnetic and electric parameters is particularly important for next generation magneto-electric devices. These interesting physical properties have led to the design of numerous novel devices. Recently, materials such as perovskite lanthanide ferrites $LnFeO_3$ have attracted much interest due to their potential applications in ultrafast lasers in magnetic devices, as well as multiferroic materials [6-10]. In these orthoferrites, the $Fe^{3+}$ ions form two magnetic sublattices where their spins are antiferromagnetically coupled. In addition, the coupling of electronic orbital states to ordered spins determines the magnetic anisotropy in magneto-electric and multiferroic materials. Understanding the magnetic interactions between the lanthanide $Ln^{3+}$-*4f* and transition metal $Fe^{3+}$-*3d* is very challenging but it greatly influences their physical properties, as demonstrated for $NdFeO_3$ [11]. Similarly, in the perovskite $ErFeO_3$ the magnetic phase transition can be explicitly explained by $Fe^{3+}(3d)$-$Er^{3+}(4f)$ exchange coupling constants [10]. Interestingly, the $SmFeO_3$ compound also belongs to a class of orthoferrites, and it has been investigated recently in several studies because of its controversial multiferroic properties [12-15]. Lee et al. [12] conducted theoretical and experimental studies of $SmFeO_3$ and showed that this phase exhibits improper ferroelectricity at room temperature and magnetization reversal at low temperatures. However, the existence of ferroelectricity in $SmFeO_3$ was refuted based on experimental investigations by Kuo et al. [15] who suggested that magneto-elastic effects lead to artificial ferroelectricity.

Based on these results, we suggest that the origin of this ambiguous ferroelectricity in $SmFeO_3$ may be explained by different synthesis conditions, which could be responsible for differences in the lossy character of the samples [12,15]. Moreover, it has been observed that the multiferroic and magneto-dielectric behavior in B-site doped ferrites $LnFeO_3$ could be achieved by using different transition metals [16,17]. Therefore, in the present study, we focused on two parameters comprising B-site doping in $SmFeO_3$, i.e., 50% iron replaced by cobalt, and the sol–gel synthesis process, which has many advantages compared with other techniques [18]. One of the most important advantages of this technique is the possibility of controlling the chemical reaction at the molecular level to obtain better crystallinity and phase formation. The nanoscale layered $SmFe_{0.5}Co_{0.5}O_3$ perovskite crystallizes in an orthorhombic structure (space group: Pnma) in a similar manner to the parent $SmFeO_3$ [12-15] and $SmCoO_3$ [19,20] phases. Recently, we reported that this perovskite exhibits a unique nanostructure built of mixed Fe-Co octahedral layers with gradual and periodic variations in the cations from Co-rich to Fe-rich layers [21]. Considering the ability of Fe/Co cations to occupy the same B-site in the perovskite $ABO_3$, the design of mixed Fe-Co perovskites could allow the exploitation of various interesting magnetic and dielectric properties. We determined an unusual and complex magnetic transition in the nanoscale layered $SmFe_{0.5}Co_{0.5}O_3$ phase as an important feature of the antiferromagnetism, which comprised an anomalous small ordered moment for $Fe^{3+}$ and $Co^{3+}$ in a similar manner to its parent phases [12,19-21]. Moreover, the nanoscale layered $SmFe_{0.5}Co_{0.5}O_3$ phase exhibited a positive magneto-dielectric effect of nearly 2.5% at a low temperature (100 K), which may be attributed to weak spin–lattice interactions.

## 2. Experimental procedure

The conventional sol–gel method was used to obtain the nanoscale layered $SmFe_{0.5}Co_{0.5}O_3$ perovskite sample [18]. Appropriate quantities of the metal nitrates and citric acid were dissolved in distilled water. Citric acid was added to the solution at a quantitative molar ratio of 1:2. The mixture was stirred at 60°C for 3 h and then left undisturbed at 100°C to allow gel formation to occur, before drying at 150°C for 12 h. The powdered compound was allowed to decompose in the air at 250°C for 24 h. The decomposed powder was then sintered several times at various temperatures ranging from 400°C to 800°C with intermediate grinding. The resulting polycrystalline powder was ground again thoroughly and pressed into pellets, which were finally sintered in the air at 950°C for 48 h. The sintered pellets were broken into small pieces and used for various measurements, before some pellets were ground again to obtain a fine powder for recording X-ray diffraction measurements in the 2θ range from 10–100°. The oxygen stoichiometry of the sample was checked by iodometric titration, which confirmed that the value for $SmFe_{0.5}Co_{0.5}O_3$ was within the accuracy limit (3 ± 0.05). The sol–gel process allows the synthesis of a single phase sample with uniform crystal morphology at a low temperature and without any secondary phase, as demonstrated by transmission electron microscopy (TEM), high resolution TEM, electron diffraction (ED), High angle annular dark field scanning TEM HAAD-STEM, and electron energy loss spectroscopy EELS studies [21]. The Physical Property Measurement System (PPMS, Quantum Design) was used to investigate the field and temperature dependences of the magnetization under different conditions. Dielectric analyses as functions of the temperature and magnetic field were conducted using the PPMS coupled with an impedance analyzer (Agilent Technologies 4284A) in the temperature range from 10 K to 300 K, frequency range from 10 kHz to 500 kHz, and an applied magnetic field up to 14 Tesla. The applied electric and magnetic fields were in the same directions in order to eliminate contributions due to the Hall Effect [22]. The electrodes on the pellets were prepared in the capacitor geometry by painting both sides of the polished pellets with silver paste.

## 3. Results and discussion

The powder diffraction pattern obtained for the nanoscale layered $SmFe_{0.5}Co_{0.5}O_3$ perovskite was indexed with orthorhombic symmetry to the *Pnma* space group [21]. The X-ray diffraction pattern is presented in Fig. 1, and the refined data and lattice parameters are given in Table 1. The temperature-dependent zero field cooled (ZFC) and field cooled (FC) magnetization results obtained for the nanoscale layered $SmFe_{0.5}Co_{0.5}O_3$ perovskite under applied fields of 50, 200, and 1000 Oe with the standard measurement protocol are shown in Fig. 2. The ZFC-FC magnetization, M(T), increased as the temperature decreased and there was a sharp magnetic transition around 310 K. The FC data increased rapidly up to 210 K, before a sudden change in the slope (a kink) of the M(T) curve occurred and the magnetization value then decreased up to 170 K. However, after further cooling, the M(T) value increased gradually and exhibited a broad maximum around 70 K. Finally, the magnetization decreased abruptly to a lower value below 40 K. However, the ZFC data obtained at 50 Oe exhibited contrasting behaviors in the temperature range considered compared with higher applied fields. There was a large divergence between ZFC-FC in the lower field (Fig. 2a), whereas the ZFC superimposed the FC data at higher applied fields (Fig. 2b–c). This complex magnetic behavior below room temperature could be explained by two inequivalent magnetic orderings at low temperatures between the cations in different magnetic sublattices, i.e., the 4f electron based Ln sublattice and 3d electron based Fe-Co sublattice, which are arranged in an antiparallel manner [17]. The ZFC magnetization curve at 50 Oe indicated antiferromagnetic ordering at 310 K ($T_N$) and a spin-reorientation transition then occurred at 210 K ($T_{SR1}$). As the temperature decreased further, the magnetization of the system changed from a negative value to a positive value with a second spin-reorientation at 22 K ($T_{SR2}$). By contrast, the FC data indicated the occurrence of a weak spin-reorientation transition below $T_N$, which diverged at 275 K and its behavior was similar to the high field M(T). For the ZFC magnetization curve at 50 Oe, the spin-reorientations according to the variations in temperature could be explained by the fact that the effective moment of the Sm sublattice rotates and aligns in opposite directions relative to the Fe-Co sublattice, thereby resulting in a negative magnetization value. In perovskite orthoferrites, these two sublattices are generally ordered magnetically at different temperatures because spin-reorientation phenomena are observed due to the strong competition between the two magnetic sublattices. This type of magnetization reversal was reported recently for the ErFe$_{0.5}$Co$_{0.5}$O$_3$ phase by Lohr et al. [17]. The Ln sublattice is ordered antiferromagnetically at relatively low temperatures, whereas the ordering temperature is higher for transition metals [10-12, 17]. It should be noted that these magnetic transitions could be the source of the magnetization due to strong coupling between the 3d and 4f electrons from the Fe-Co and Ln sublattices. Therefore, the magnetic kink at 210 K and the abrupt decrease in the M(T) value below 40 K could be explained by spin-reorientation.

Figure 3 shows the field-dependent magnetization behavior M(H) at 2, 100, and 350 K for the nanoscale layered SmFe$_{0.5}$Co$_{0.5}$O$_3$ in order to illustrate the different types of magnetic interactions. The measurements were performed after zero field cooling the sample from room temperature to the desired temperature and then obtaining a full hysteresis loop (including the virgin curve). The system exhibited a hysteresis loop with large values for the remanent magnetization (0.012 μ$_B$/f.u.) and coercive field (6 kOe) at 100 K. Importantly, the magnetization value did not indicate any saturation even when the applied field strength was 5 Tesla, but instead the value increased in an almost linear manner with the applied field, and the maximum magnetization value was 0.066 μ$_B$/f.u., which is similar to that reported for the parent SmFeO$_3$ perovskite [12]. The M(H) behavior at 2 K was similar to that at 100 K. However, the corresponding remanent magnetization and coercive field values were lower (0.005 μ$_B$/f.u. and 2 kOe) compared with the expected values for magnetic perovskites, possibly due to large domain wall pinning of the magnetic spins at 100 K. Finally, at 350 K, the M(H) data exhibited linear behavior in an analogous manner to the paramagnetic system. Hence, the behavior of the magnetic hysteresis loops below $T_N$ clearly indicated that the nanoscale layered SmFe$_{0.5}$Co$_{0.5}$O$_3$ behaved as a canted antiferromagnet, where weak ferromagnetism remained due to canting of the magnetic spins. The magnetic ordering at 310 K could have been due to the canted nature of the magnetic spins, which are antiferromagnetically aligned [23]. This feature is well known for orthoferrites where the weak ferromagnetism is due to non-collinear antiferromagnetism in the Fe sublattice at higher temperatures [11-17]. The weak ferromagnetism in these orthoferrites could also be attributed to the asymmetric exchange or Dzyaloshinskii–Moriya interaction, which results in canted antiferromagnetic ordering, thereby implying that the nonlinear motions of each sublattice spin along the net magnetization direction do not cancel perfectly [17, 24]. This type of interaction is well established for the perovskite orthoferrite [10-20, 23, 24].

To investigate the magneto-dielectric correlations, we studied the dependences of the dielectric constant and dielectric loss on the temperature and magnetic field for the nanoscale layered SmFe$_{0.5}$Co$_{0.5}$O$_3$ perovskite at various frequencies. Figure 4 shows the dependences of the real part of the dielectric constant ($\varepsilon'$) and dielectric loss (tan$\delta$) on the temperature at various frequencies. The $\varepsilon'$ value increased as the temperature increased with a rapid jump around 100 K, where further increases in temperature did not result in a maximum $\varepsilon'$ value as usually found for conventional dielectrics, but instead the $\varepsilon'$ value increased continuously with the temperature. This type of behavior can be explained by dipolar type relaxation coupled with variable type hopping conduction of the charge carrier, which is a common phenomenon for semiconductors [22, 25, 26]. The localized charge carrier usually hops between spatially fluctuating lattice potentials to produce conductivity and the dipolar effect.

This type of dielectric behavior has been reported for polycrystalline Fe-doped perovskites [26]. It should be noted that the magnitude of the dielectric constant for this perovskite system decreased as the frequency increased, which indicates that the nanoscale layered $SmFe_{0.5}Co_{0.5}O_3$ perovskite exhibited relaxor-like behavior. Prominent tanδ peaks (Fig. 4b) were observed in the region where the behavior of the dielectric constant changed rapidly. The shift of the tanδ peaks toward higher temperatures as the frequency increased indicated thermally activated relaxation. Due to the presence of dc conductivity in this perovskite (semiconducting type), tanδ increased with the temperature, i.e., the loss (tanδ) near room temperature was double compared with the peak value (Fig. 4b).

According to previous studies, it is clear that magnetic and dielectric transitions occur below room temperature, which could lead to correlations between the magnetic and dielectric properties. Hence, we conducted isothermal magneto-dielectric studies at different temperatures. Figure 5(a) shows the dependence of the dielectric constant on the magnetic field for the nanoscale layered $SmFe_{0.5}Co_{0.5}O_3$ perovskite at four different temperatures and at a high frequency of 100 kHz [22]. A positive magneto-dielectric value of 2.5% was obtained at 100 K, which was calculated as $\Delta\varepsilon = [\{\varepsilon_H – \varepsilon_0\}/\varepsilon_0] \times 100$, where $\varepsilon_H$ and $\varepsilon_0$ are the dielectric constant values in the presence and absence of magnetic fields, respectively [22, 25]. The values were significantly smaller above and below this temperature, although magneto-dielectric coupling was prominent above 100 K. Thus, in order to understand the source of the behavior observed at high temperatures, we conducted temperature-dependent dielectric analyses in the presence and absence of a magnetic field. Figures 5(b) and 5(c) show the dependences of $\varepsilon'$ and loss (tanδ) on the temperature of the system under different magnetic fields (frequency at 100 kHz). The dielectric constant value varied from 10 to 170 in the temperature range of 10–300 K. At low temperatures (< 90 K), the dielectric constant values were independent of the frequency and temperature. Around 100 K, a sudden increase in the dielectric constant value was observed. The rapid increase in the dielectric constant values above 100 K could be explained by numerous factors that may be intrinsic and/or extrinsic to the system [22, 25-30], and thus we analyzed the data in detail to confirm the actual effect. The rapid increase in the $\varepsilon'$ value within a small temperature interval may have occurred due to the onset of ferroelectric ordering where the dipoles are subjected to a double well energy barrier [29]. This effect is intrinsic to the system and the $\varepsilon'$ value follows the Curie–Weiss behavior beyond the ferroelectric ordering. However, for the nanoscale layered $SmFe_{0.5}Co_{0.5}O_3$ perovskite, we observed no peak in the $\varepsilon'$ value. Hence, the combined effect of the grain boundary and space charge effect may have been responsible for the behavior observed above 100 K due to the semiconducting nature of the nanoscale layered $SmFe_{0.5}Co_{0.5}O_3$ perovskite associated with contact effects. By contrast, the low temperature magneto-dielectric coupling was only due to the intrinsic spin–lattice interaction, which led to a positive magneto-capacitance effect with a maximum value around 100 K. It should be noted that it has been reported that the positive value for the magneto-dielectric constant originates from the spin–lattice interaction, which is an intrinsic property of the system [22, 25, 30]. In addition, the magnetic field-dependent dielectric loss (tan$\delta$) at 100 kHz (inset in Fig. 5c) exhibited a similar trend to the shift in the $\varepsilon'$ value, as expected for an intrinsic magneto-dielectric effect [22].

## 4. Conclusions

In this study, we analyzed the low temperature magnetic and dielectric behavior as well as the magneto-dielectric coupling of a mixed iron-cobalt based nanoscale layered $SmFe_{0.5}Co_{0.5}O_3$ perovskite. The complex nature of the magnetic behavior was reflected by multiple transitions in the low temperature regions. This complex behavior can be attributed to the different ordering temperatures of the magnetic sublattices and interactions between the 3d and 4f sublattices. The magnetic interactions were mainly antiferromagnetic in nature. However, the antisymmetric or Dzyaloshinskii–Moriya interaction may be considered the source of the weak ferromagnetism in the nanoscale layered $SmFe_{0.5}Co_{0.5}O_3$. Dielectric measurements demonstrated the relaxor-like behavior at low temperatures. More importantly, we observed an intrinsic magneto-dielectric effect of ~ 2.5% around 100 K, which suggests active coupling between the ordering in the magnetic and electric parameters, and this is of crucial importance for the next generation of magneto-electric devices. We hope that these results facilitate the exploration of materials with large magneto-dielectric coupling at room temperature.

**Acknowledgments**

The authors would like to thank the Science and Engineering Research Board (SERB), India for financial support through project grant # EMR/2016/000083, and Prof. B. Raveau for his valuable comments and suggestions.

# References

[1] A.V. Kimel, A. Kirilyuk, P.A Usachev, R.V Pisarev, A.M Balbashov, T. Rasing, Nat. 435 (2005) 655–657.

[2] M. Fiebig, J. Phys. D 38 (2005) R123–R152.

[3] W. Eerenstein, M. Wiora, J. L. Prieto, J. F. Scott, N. D. Mathur, Nat. Mater. 6 (2007) 348–351.

[4] Y. Tokunaga, S. Iguchi, T. Arima, Y. Tokura, Phys. Rev. Lett. 101 (2008) 097205.

[5] Y. Tokunaga, N. Furukawa, H. Sakai, Y. Taguchi, T. Arima, Y. Tokura, Nat. Mater. 8 (2009) 558–562.

[6] R.V. Mikhaylovskiy, E. Hendry, A. Secchi, J.H. Mentink, M. Eckstein, A. Wu, R.V. Pisarev, V.V. Kruglyak, M.I. Katsnelson, T. Rasing, A.V. Kimel, Nat. Commun. 6 (2015) 8190.

[7] R. V. Mikhaylovskiy, T. J. Huisman, R. V. Pisarev, T. Rasing, A. V. Kimel, Phys. Rev. Lett. 118 (2017) 017205.

[8] J.A. de Jong, A.V. Kimel, R.V. Pisarev, A. Kirilyuk, T. Rasing, Phys. Rev. B 84 (2011) 104421.

[9] T. F. Nova, A. Cartella, A. Cantaluppi, M. Först, D. Bossini, R. V. Mikhaylovskiy, A. V. Kimel, R. Merlin, A. Cavalleri, Nat. Phys. 13 (2017) 132–136.

[10] X. Li, M. Bamba, N. Yuan, Q. Zhang, Y. Zhao, M. Xiang, K. Xu, Z. Jin, W. Ren, G. Ma, S. Cao, D. Turchinovich, J. Kono, Science 361 (2018) 794–797.

[11] S.J. Yuan, W. Ren, F. Hong, Y.B. Wang, J.C. Zhang, L. Bellaiche, S.X. Cao, G. Cao, Phys. Rev. B 87 (2013) 184405.

[12] J.H. Lee, Y. K. Jeong, J. H. Park, M.-A. Oak, H. M. Jang, J. Y. Son, J. F. Scott, Phys. Rev. Lett. 107 (2011) 117201.

[13] R. D. Johnson, N. Terada, P. G. Radaelli, Phys. Rev. Lett. 108 (2012) 219701.

[14] J.H. Lee, Y. K. Jeong, J. H. Park, M.-A. Oak, H. M. Jang, J. Y. Son, J. F. Scott, Phys. Rev. Lett. 108 (2012) 219702.

[15] C.Y. Kuo, Y. Drees, M. T. Fernández-Díaz, L. Zhao, L. Vasylechko, D. Sheptyakov, A. M. T. Bell, T. W. Pi, H.-J. Lin, M.-K. Wu, E. Pellegrin, S. M. Valvidares, Z. W. Li, P. Adler, A. Todorova, R. Küchler, A. Steppke, L. H. Tjeng, Z. Hu, A. C. Komarek, Phys. Rev. Lett. 113 (2014) 217203.

[16] G. Kotnana, S. N. Jammalamadaka, J. Mag. Mag. Mater. 418 (2016) 81–85.

[17] J. Lohr, F. Pomiro, V. Pomjakushin, J. A. Alonso, R. E. Carbonio, R. D. Sánchez, Phys. Rev. B 98 (2018) 134405.

[18] V. Solanki, S. Das, S. Kumar, M. M. Seikh, B. Raveau, A. K. Kundu, J. Sol-Gel Sci. Technol. 82 (2017) 536.

[19] N. B. Ivanova, N. V. Kazak, C. R. Michel, A. D. Balaev, S. G. Ovchinnikov, Physics of the Solid State 49 (2007) 2126–2131.

[20] M. Tachibana, T. Yoshida, H. Kawaji, T. Atake, E. T. Muromachi, Phys. Rev. B 77 (2008) 094402.

[21] N. E. Mordvinova, A. Shukla, A. K. Kundu, O. I. Lebedev, M. M. Seikh, V. Caignaert, B. Raveau, arxiv (http://arxiv.org/abs/1811.05221).

[22] H. M. Rai, S. K. Saxena, V. Mishra, R. Kumar, P. R. Sagdeo, J. App. Phys., 122, (2017) 054103.

[23] D. V. Karpinsky, I. O. Troyanchuk, V. V. Sikolenko, J. Phys.: Condens. Matter 17 (2005) 7219.

[24] J. Lu, X. Li, H. Y. Hwang, B. K. Ofori-Okai, T. Kurihara, T. Suemoto, K. A. Nelson, Phys. Rev. Lett., 118 (2017) 207204

[25] G. Catalan, Appl. Phys. Lett., 88 (2006) 102902.

[26] A. K. Kundu, V. K. Jha, M. M. Seikh, R. Chatterjee, R. Mahendiran, J. Phys.: Condens. Matter 24 (2012) 255902.

[27] T. Moriya, Phys. Rev. Lett. 4 (1960) 228.

[28] L. E. Cross, Ferroelectrics 76 (1987) 241–267.

[29] P. Lunkenheimer, V. Bobnar, A. V. Pronin, A. I. Ritus, A. A. Volkov, A. Loidl, Phys. Rev. B 66 (2002) 052105.

[30] A. K. Kundu, R. Ranjith, V. Pralong, V. Caignaert, B. Raveau, J. Mater. Chem., 18 (2008) 4280–4285.

**Table 1.** Lattice parameters for the nanoscale layered $SmFe_{0.5}Co_{0.5}O_3$ perovskite. $a$, $b$, and $c$ are the lattice parameters, and $R_b$ and $R_f$ are the Bragg factor and fit factor, respectively.

| Perovskite | $SmFe_{0.5}Co_{0.5}O_3$ |
|---|---|
| Space group | *Pnma* |
| $a$ (Å) | 5.484 (6) |
| $b$ (Å) | 7.611 (3) |
| $c$ (Å) | 5.348 (4) |
| V (Å$^3$) | 223.410 |
| α, β, γ | 90°, 90°, 90° |
| Bond length | Sm-O1: 2.305 Å |
| | Sm-O1: 2.444 Å |
| | Sm-O2: 2.444 Å |
| | Sm-O2: 2.305 Å |
| | Fe/Co-O1: 1.968 Å |

|  | Fe/Co-O1: 1.972 Å |
|  | Fe/Co-O2: 1.974 Å |
| Bond angle | <Fe/Co-O1-Fe/Co>: 159.7° |
|  | <Fe/Co-O2-Fe/Co>: 157.3° |
| $R_b$ (%) | 11.3 |
| $R_f$ (%) | 19.9 |
| $\chi$-factor | 2.87 |

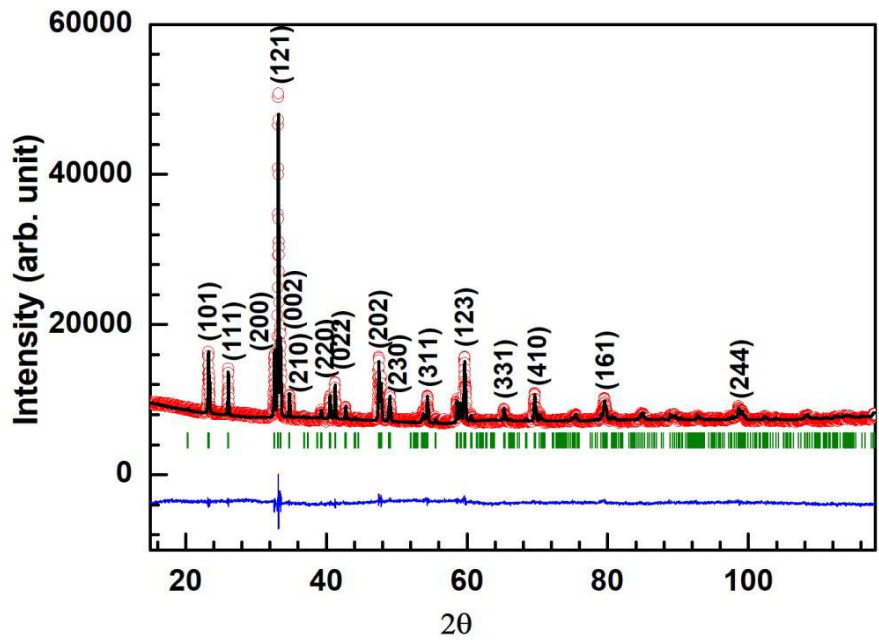

**Fig. 1.** Rietveld analysis of the XRD pattern obtained for nanoscale layered $SmFe_{0.5}Co_{0.5}O_3$. Open symbols are experimental data and the solid and vertical lines represent the difference curve and Bragg positions, respectively.

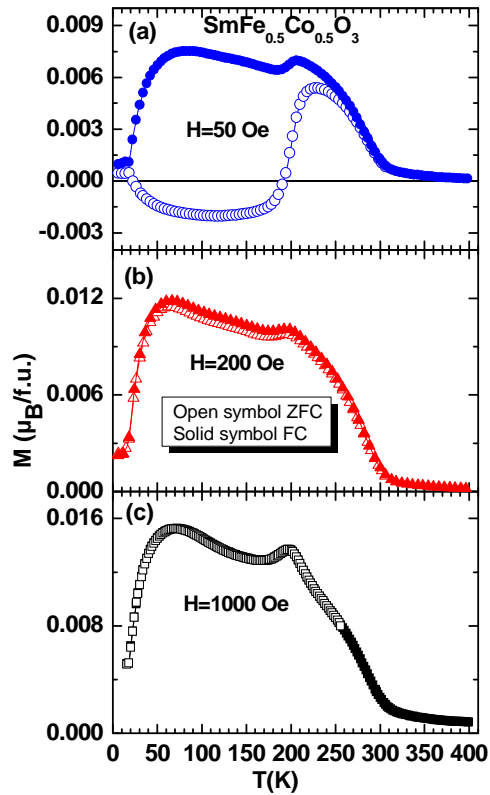

**Fig. 2.** Temperature-dependent ZFC (open symbol) and FC (solid symbol) magnetization, M, for nanoscale layered $SmFe_{0.5}Co_{0.5}O_3$ under the fields of (a) H = 50 Oe, (b) H = 200 Oe, and (c) H = 1000 Oe.

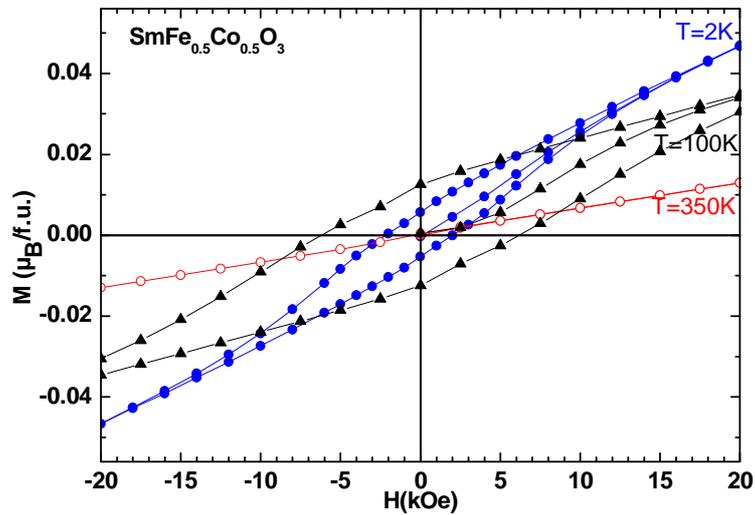

**Fig. 3.** Field-dependent isothermal magnetic hysteresis, M(H), curves obtained at three different temperatures for nanoscale layered $SmFe_{0.5}Co_{0.5}O_3$.

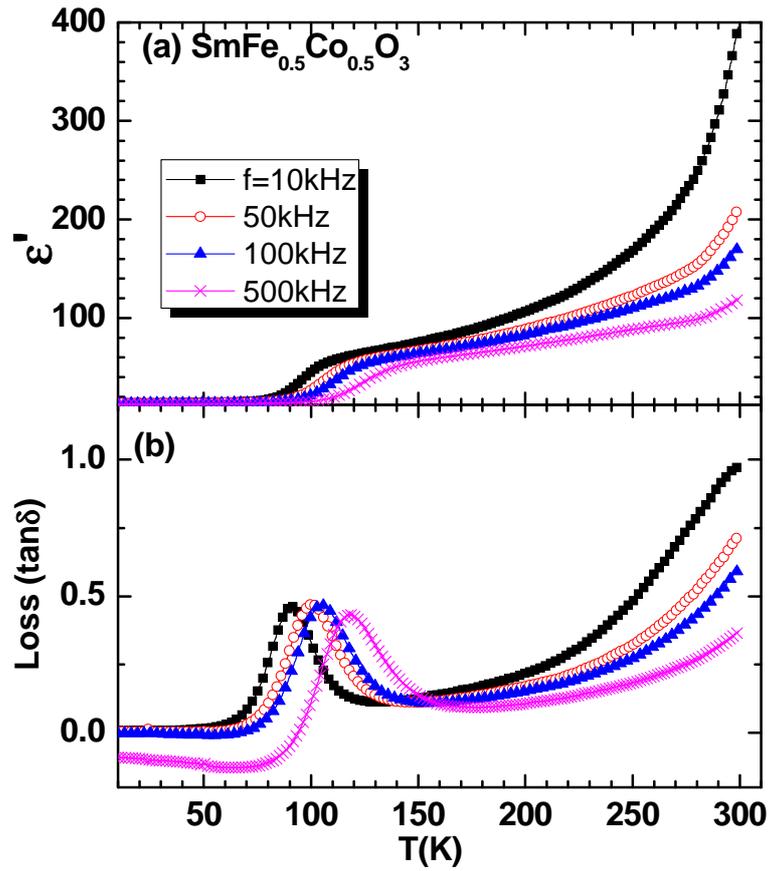

**Fig. 4.** Temperature dependence of the (a) dielectric constant and (b) dielectric loss for nanoscale layered $SmFe_{0.5}Co_{0.5}O_3$ at different frequencies.

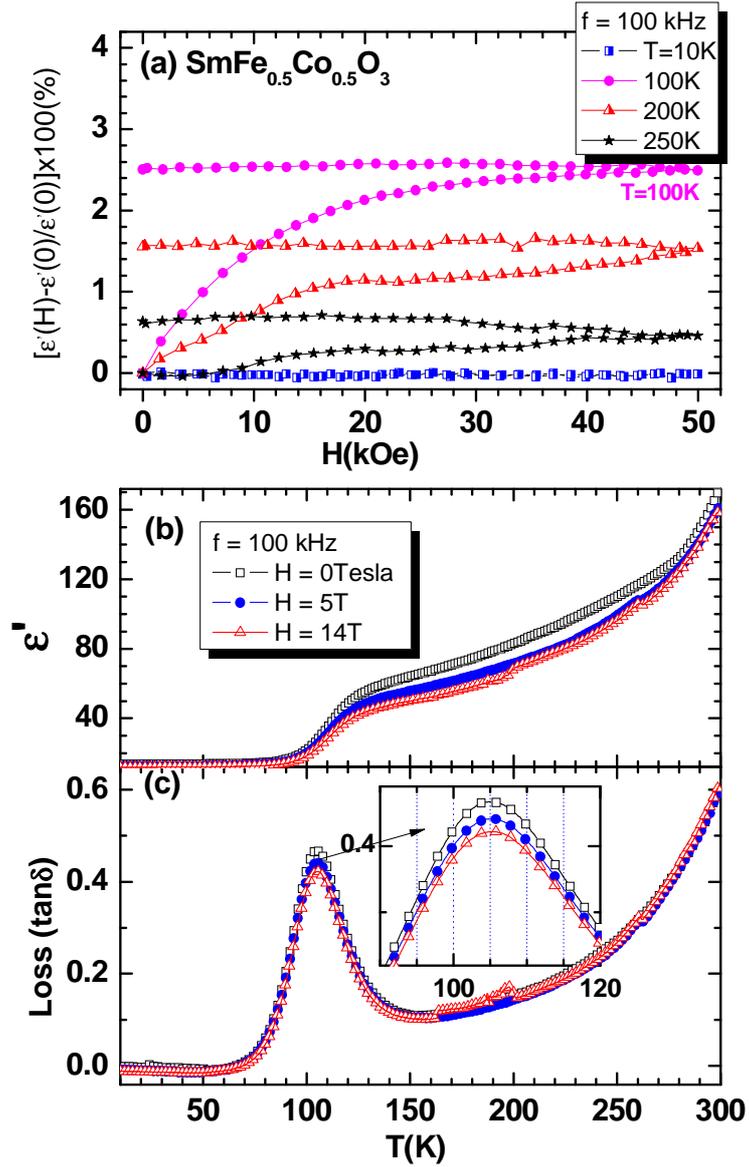

**Fig. 5.** (a) Field-dependent isothermal dielectric constant at four different temperatures and (b–c) temperature-dependent dielectric constant and loss in three different fields for nanoscale layered $SmFe_{0.5}Co_{0.5}O_3$. The inset shows the magnified dielectric loss peak.